\documentclass[11pt]{elsarticle}
\usepackage{amsmath,amssymb,graphicx,float,latexsym,psfrag,subfigure,color}

\newcommand{\pd}[2]{\frac{\partial #1 }{\partial #2}}

\newcommand{\Pdc}[3]{\frac{\partial^{2} #1 }{\partial #2 \partial #3}}

\newcommand{\lpf}[1]{\overline{#1}}
\newcommand{\tlpf}[1]{\widehat{#1}}
\newcommand{\beq}{\begin{equation}}
\newcommand{\eeq}{\end{equation}}
\newcommand{\N}{\mathcal{N}}

\newcommand{\R}{\mathcal{R}}
\newcommand{\U}{\mathcal{U}}

\newcommand{\G}{\mathcal{G}}
\newcommand{\M}{\mathcal{M}}

\newcommand{\fDelta}{{\lpf{\Delta}}}
\newcommand{\ffDelta}{\tlpf{\lpf{\Delta}}}
\newcommand{\ff}[1]{\tlpf{\lpf{#1}}}

\newcommand{\etal}{\emph{et al.}}

\begin{document}
\begin{frontmatter}

  \title{An alternative derivation of the Germano identity as the residual of the LES equation}
  
\author{Siavash Toosi\fnref{stfootnote}}\ead{stoosi@umd.edu}
\author{Johan Larsson}\ead{jola@umd.edu}
\address{Department of Mechanical Engineering, University of Maryland, College Park, MD 20742, USA}
\fntext[stfootnote]{Current address:
Linn\'{e} FLOW Centre,
KTH Mechanics, SE-10044 Stockholm, Sweden}

\end{frontmatter}

\section{Introduction}

The Germano identity~\cite{germano:92}
and the resulting dynamic procedure~\cite{germano:91,lilly:92}
to compute subgrid model coefficients
have been among the most successful and popular developments
in large eddy simulation (LES).
The original rationale for the dynamic procedure
was that the same subgrid model should be applicable with the same model coefficient at two different coarse-graining levels (or filter levels), which was later interpreted as an argument based on scale-invariance (cf.~\cite{meneveau:00}),
a property that is expected of turbulence in the inertial subrange.
The rationale based on scale-invariance was first
questioned by Jimenez and Moser~\cite{jimenez:00} and later by
Pope~\cite{pope:04}, partly based on the fact that the dynamic procedure works well at low Reynolds numbers (transitional flow, near-wall behavior, etc.) where scale-invariance does not hold.
Jimenez and Moser~\cite{jimenez:00} argued that 
the success of the dynamic procedure is probably due
to the balance between the production of Leonard stresses
and the dissipation rate resulting from the application of the dynamic procedure.
Pope~\cite{pope:04}, on the other hand,
argued that the reason for success is that the dynamic procedure 
minimizes the sensitivity of the total (resolved plus modeled) Reynolds stresses to the coarse-graining level.
This general argument was later used by Meneveau~\cite{meneveau:12} as well.
The current consensus understanding of why the dynamic procedure works
seems to be a combination of 
these arguments,
the strength of which is that they require no specific assumptions about the 
characteristics of the flow 
(e.g., whether it satisfies the scale-similarity hypothesis)
and to some degree about its nature
(e.g., whether it is turbulent or not).

The objective of this Note is to present an alternative derivation of the Germano identity and its error
which provides a subtly different argument for why the dynamic procedure works.
While the previous arguments rest on recognizing the importance of 
the Reynolds stress or the dissipation rate,
the present derivation instead follows the path of deriving the residual 
(in the sense of numerical analysis)
of the LES equation.
The residual is of central importance in the field of numerical analysis
since it is the source of errors; 
therefore, the present derivation shows the connection between 
the error in the Germano identity and the source of error in LES based on the governing equation alone, with no physical insight required.
The present derivation does not contradict the prior arguments by 
Jimenez and Moser~\cite{jimenez:00} or by 
Pope~\cite{pope:04} and Meneveau~\cite{meneveau:12} in any way; 
rather, it is offered here as a complement to the prior explanations.

\section{LES equations}

The Navier-Stokes equation for 
an incompressible and constant viscosity fluid is
\beq \nonumber
\pd{{\mathcal{U}}_i}{t}
+
\pd{{\mathcal{U}}_i {\mathcal{U}}_j}{x_j}
+\frac{1}{\rho}
\pd{{\mathcal{P}}}{x_i}
-\nu
\Pdc{ {\mathcal{U}}_i }{x_j}{x_j}
=0
\, ,
\eeq
or in short notation 
$\mathcal{N}({\mathcal{U}_i}) = 0$,
where $\rho$ and $\nu$ are the density and viscosity 
(both assumed constant here)
and
$\mathcal{U}_i$ and $\mathcal{P}$
are the exact velocity and pressure fields
(corresponding 
to a perfect direct numerical simulation, DNS).
When coarse-grained or implicitly filtered 
to a resolution length scale (or filter width) $\fDelta$
the equation becomes
(assuming that filtering and differentiation commute)
\beq \label{eq:exact_filtered}
\pd{\lpf{\mathcal{U}}_i}{t}
+
\pd{\lpf{\mathcal{U}}_i \lpf{\mathcal{U}}_j}{x_j}
+\frac{1}{\rho}
\pd{\lpf{\mathcal{P}}}{x_i}
-\nu
\Pdc{ \lpf{\mathcal{U}}_i }{x_j}{x_j}
+
\pd{\tau_{ij,\fDelta}^{{\rm exact}}}{x_j}
=0
\, ,
\eeq
or $\mathcal{N}_\fDelta^{\rm exact}(\lpf{\mathcal{U}_i}) = 0$
where $\lpf{\mathcal{U}_i}$ and $\lpf{\mathcal{P}}$
are coarse-grained representations of the exact fields
and $\tau_{ij,\fDelta}^{{\rm exact}} = \lpf{\mathcal{U}_i \mathcal{U}_j}  - \lpf{\mathcal{U}}_i \lpf{\mathcal{U}}_j$
is the exact subgrid scale (SGS) stress tensor.
An interesting property of $\tau_{ij}^{{\rm exact}}$ is that 
it satisfies the Germano identity~\cite{germano:92}
\beq \label{eq:GI}
\widehat{\tau_{ij,\fDelta}^{{\rm exact}} }
-
\tau_{ij,\ffDelta}^{{\rm exact}} 
=
\widehat{ \lpf{\mathcal{U}}_i \lpf{\mathcal{U}}_j }
-
\ff{\mathcal{U}}_i \ff{\mathcal{U}}_j
\,,
\eeq
where $\widehat{\cdot}$ is a test filtering operation 
of width $\widehat{\Delta}>\fDelta$,
$\ff{\cdot}$ is the result of consecutive application of 
filters ${\fDelta}$ and $\widehat{\Delta}$, 
and $\tau_{ij,\ffDelta}^{{\rm exact}} = \widehat{ \lpf{\mathcal{U}_i \mathcal{U}_j}}  - \ff{\mathcal{U}}_i \ff{\mathcal{U}}_j$.
This identity provides a ``self-consistency condition''~\cite{meneveau:12}
that also applies to
Eqn.~(\ref{eq:exact_filtered})
at filter levels $\fDelta$ and $\ffDelta$.


Approximating the exact SGS stress tensor using a model leads to the LES equation in differential form 
(i.e., without numerical errors),
where we intensionally exclude the numerical errors in order 
to isolate the effect of modeling errors in the equation,
and to be faithful to many of the developments 
in the LES literature.
The LES equations at two different filter levels
$\fDelta$ 
(say, original)
and $\ffDelta$
(test filtered)
are
\begin{align}
\label{eq:les}
\pd{\lpf{u}_i}{t}
+
\pd{\lpf{u}_i \lpf{u}_j}{x_j}
+\frac{1}{\rho}
\pd{\lpf{p}}{x_i}
-\nu
\Pdc{ \lpf{u}_i }{x_j}{x_j}
+
\pd{\tau_{ij,\fDelta}^{{\rm model}} (\lpf{u}_k)}{x_j}
&=0
\,,
\\
\label{eq:les_testlevel}
\pd{\ff{v}_i}{t}
+
\pd{\ff{v}_i \ff{v}_j}{x_j}
+\frac{1}{\rho}
\pd{\ff{q}}{x_i}
-\nu
\Pdc{ \ff{v}_i }{x_j}{x_j}
+
\pd{\tau_{ij, \ffDelta}^{{\rm model}} (\ff{v}_k)}{x_j}
&=0
\, ,
\end{align}
where $(\lpf{u}_i, \lpf{p})$ 
and
$(\ff{v}_i, \ff{q})$ are the solutions
at the respective filter levels.
These equations are referred to as
$\N_\fDelta^{\rm model} (\lpf{u}_i) =0$
and
$\N_{\ffDelta}^{\rm model} (\ff{v}_i) =0$,
respectively.


The principle of the dynamic procedure~\cite{germano:91} 
is that any approximate 
model
should satisfy, 
as well as possible, the Germano identity. 
It therefore aims to minimize the error
\beq \label{eq:GIE}
\mathcal{G}_{ij}
=
\underbrace{
\widehat{ \lpf{{u}}_i \lpf{{u}}_j }
-
\ff{{u}}_i \ff{{u}}_j
}_{\mathcal{L}_{ij}}
-
\underbrace{
\left[
\widehat{\tau_{ij,\fDelta}^{{\rm model}}(\lpf{u}_k) }
-
\tau_{ij,\ffDelta}^{{\rm model}} (\ff{u}_k) 
\right]
}_{\mathcal{M}_{ij}}
\, ,
\eeq
in a least squares sense~\cite{lilly:92}.
Here, $\G_{ij}$ is the Germano identity error (GIE),
$\mathcal{L}_{ij}$ is the Leonard or resolved stress,
and $\M_{ij}$ is the modeled stress~\cite[cf.][]{pope:00}.

\section{Residual due to modeling and its connection to the modeling error}
 
The residual of an inexact equation $\N_{\rm approx}(u_{\rm approx})=0$
is the misfit when evaluating the inexact equation for the exact solution, i.e.,
$\N_{\rm approx}(u_{\rm exact})$ in this example.
The importance of the residual is made clear by the Taylor expansion
\beq \nonumber
\N_{\rm approx}(u_{\rm exact}) \approx
\underbrace{ \N_{\rm approx}(u_{\rm approx}) }_{=0}
+ \pd{ \N_{\rm approx} }{ u }
\underbrace{ \left( u_{\rm approx} - u_{\rm exact} \right) }_{\rm error}
\,,
\eeq
which shows how the residual is the source of error in the solution for linearized dynamics.
We therefore want to find the residual of the LES equation (\ref{eq:les}).
This residual is $\N_\fDelta^{\rm model} (\lpf{\mathcal{U}}_i)$
for filter level $\fDelta$
and $\N_{\ffDelta}^{\rm model} (\ff{\mathcal{U}}_i)$
for filter level $\ffDelta$,
where we must use the coarse-grained representations of the exact fields 
(clearly the full field $\mathcal{U}_i$ is consistent only with the DNS equation, 
not the coarse-grained ones containing $\tau_{ij}$).
Therefore, we can write
\beq \nonumber
\begin{aligned}
\R_{\fDelta} \equiv
\N_{\fDelta}^{\rm model} (\lpf{\U}_i) 
&=
\underbrace{
\pd{\lpf{\mathcal{U}}_i}{t}
+
\pd{\lpf{\mathcal{U}}_i \lpf{\mathcal{U}}_j}{x_j}
+\frac{1}{\rho}
\pd{\lpf{\mathcal{P}}}{x_i}
-\nu
\Pdc{ \lpf{\mathcal{U}}_i }{x_j}{x_j}
}_{
= - \partial \tau_{ij,\fDelta}^{{\rm exact}} / \partial x_j
}
+
\pd{\tau_{ij, \fDelta}^{{\rm model}}(\lpf{\mathcal{U}}_k )}{x_j} 
\\
&=
\pd{}{x_j}  
\left[ 
\tau_{ij,\fDelta}^{{\rm model}}(\lpf{\mathcal{U}}_k )
-
\tau_{ij,\fDelta}^{{\rm exact}}
\right]
,
\end{aligned}
\eeq
where Eqn.~(\ref{eq:exact_filtered})
is used to replace the terms
by the divergence of $\tau_{ij,\fDelta}^{{\rm exact}}$. 
Similarly, we have
\beq  \label{eq:real-resiual-2}
\R_{\ffDelta} \equiv
\N_{\ffDelta}^{\rm model} (\ff{\U}_i) 
=
\pd{}{x_j}  
\left[ 
\tau_{ij,\ffDelta}^{{\rm model}}(\ff{\mathcal{U}}_k )
-
\tau_{ij,\ff{\Delta}}^{{\rm exact}}
\right]
.
\eeq

The exact solution is unknown and must therefore be approximated.
In the area of numerical analysis, 
the residual is often approximated by evaluating the numerical operators 
on a finer representation of the solution~\cite[cf.][]{fidkowski:11}.
This approach, however, does not work for 
LES equations
with modeling of the discarded scales,
because 
obtaining an exact (or sufficiently more accurate) LES equation
requires a more accurate $\tau_{ij}$,
which in turn requires the estimation of $\lpf{u_i u_j}$
from only the LES solution $\lpf{u}_i$, which
is impossible due to the limited spectral content of the filtered solution and the 
projection errors~\cite[cf.][]{carati:01:jfm}. 
The solution is to use the $\N^{\rm model}$ operator
(i.e., the direct approach, as in Eqn.~\ref{eq:real-resiual-2},
to avoid estimation of $\tau_{ij}^{\rm exact}$)
and to compute the residual at a coarser filter level, 
specifically the test filter level $\ffDelta$,
such that the test filtered solution $\ff{u}_i$
can be used in place of $\ff{\U}_i$
to compute the approximate residual.
We should note that approximating $\ff{\U}_i$ by $\ff{u}_i$
is only done for the purpose of estimating the residual
(very similar to the use of the numerical solution for estimating the 
truncation errors in numerical analysis),
and is assumed to be a much weaker approximation than saying that
$\ff{u}_i$ is an accurate representation of $\ff{\U}_i$ in general.

With this approximation,
the residual at the test-filter level $\ffDelta$ is
\beq \label{eq:res1}
\R_{\ffDelta}
\approx
\pd{\ff{u}_i}{t}
+
\pd{\ff{u}_i \ff{u}_j}{x_j}
+\frac{1}{\rho}
\pd{\ff{p}}{x_i}
-\nu
\Pdc{ \ff{u}_i }{x_j}{x_j}
+
\pd{\tau_{ij,\ffDelta}^{{\rm model}} (\ff{u}_k)}{x_j}
\, .
\eeq
%
This equation
can be directly computed to estimate $\R_{\ffDelta}$;
however, quite interestingly, it can be simplified by test-filtering the LES equation~(\ref{eq:les})
and subtracting it from Eqn.~(\ref{eq:res1}), which yields
(assuming that filtering and differentiation commute)
\beq \label{eqn:residual_gie}
\begin{aligned}
\R_{\ffDelta}
&\approx
\pd{}{x_j}
\left[
\ff{u}_i \ff{u}_j
+
\tau_{ij,\ffDelta}^{{\rm model}} (\ff{u}_k)
-
\tlpf{ \lpf{u}_i \lpf{u}_j}
-
\tlpf{ \tau_{ij,\fDelta}^{{\rm model}} (\lpf{u}_k) } 
\right]
\\
&=
\pd{}{x_j}
\left[
\ff{u}_i \ff{u}_j
-
\tlpf{ \lpf{u}_i \lpf{u}_j}
+
\tau_{ij,\ffDelta}^{{\rm model}} (\ff{u}_k)
-
\tlpf{ \tau_{ij,\fDelta}^{{\rm model}} (\lpf{u}_k) } 
\right]
\\ &=
-
\pd{}{x_j}
\left[
\mathcal{L}_{ij} - \mathcal{M}_{ij}
\right]
\,,
\end{aligned}
\eeq
where $\mathcal{L}_{ij}$ and $\mathcal{M}_{ij}$
are the familiar Leonard (resolved) and modeled stress terms from Eqn.~(\ref{eq:GIE}).
In other words, we have
\beq \label{eq:best}
-
\pd{}{x_j}
\left[
\mathcal{L}_{ij} - \mathcal{M}_{ij}
\right]
\approx
\R_{\ffDelta}
=
\pd{}{x_j}  
\left[ 
\tau_{ij,\ffDelta}^{{\rm model}}(\ff{\mathcal{U}}_k )
-
\tau_{ij,\ff{\Delta}}^{{\rm exact}}
\right]
.
\eeq

Therefore, the residual
$\mathcal{R}_{\ffDelta}$
of the LES equation at the test-filter level 
is approximately equal to the divergence of the error in the Germano identity
$\mathcal{L}_{ij} - \mathcal{M}_{ij}$,
and the tensor
$\mathcal{L}_{ij}-\M_{ij}$
directly estimates the modeling error 
$\tau_{ij}^{{\rm model}}-\tau_{ij}^{{\rm exact}}$.
Minimizing this Germano identity error (GIE) thus directly minimizes the 
modeling errors and the
residual that is the source of errors
in the (test-filtered) LES equation.

\section{Concluding remarks}

This Note illustrates the close connection between the residual of the test-filtered LES evolution equation
and the error in the Germano identity
(the GIE, generally written as $\mathcal{L}_{ij} - \mathcal{M}_{ij}$ in most texts,~\cite[cf.][]{pope:00}).
Equation~(\ref{eq:best}) also
shows that the GIE
approximates the difference between the modeled SGS stress tensor and the exact one given the exact solution.
This 
can explain why the dynamic procedure 
is successful at distinguishing between laminar, transitional,
and turbulent flows,
and why it is capable of
recovering the correct near-wall behavior 
of the eddy viscosity
at the vicinity of solid walls:
the exact SGS stress tensor
$\tau_{ij}^{\rm exact}$ has all these properties built in~\cite[cf.][]{vreman:04,silvis:17},
and by minimizing the difference between 
$\tau_{ij}^{\rm exact}$ and $\tau_{ij}^{\rm model}$,
the SGS model should inherit (to the largest degree possible given the chosen model form) those characteristics.

The main purpose of this Note is to complement prior interpretations of why the dynamic procedure works, 
and to serve as a connection between the fields of LES and numerical analysis.
There is a great body of work in the numerical analysis literature that utilizes the residual
to, for example, 
produce error estimates and to optimally adapt the computational grid.
The connection between the GIE and the residual suggests that the
dynamic procedure in a sense uses the same residual to improve the solution 
by optimally choosing the model parameter(s).
The implication of such a connection is that many of the more advanced
techniques that are currently used in residual minimization
(weighting the residual by the adjoints, for instance)
could (and maybe should?) be used in the dynamic procedure as well.
Furthermore, the present derivation implies
that one should be minimizing the volume integral of the residual (i.e., the GIE)
as the more meaningful and more optimal approach of reducing the errors
(optimally the GIE should be weighted by some adjoint field),
as done by Ghosal~\etal~\cite{ghosal:95},
and shows clearly that we should indeed be minimizing the divergence of the GIE rather than error itself, 
as done by Morinishi and Vasilyev~\cite{morinishi:01}.

Finally, the implications of the findings of this Note extend
to uncertainty quantification (UQ) 
and output-based grid/filter adaptation in LES,
both of which require an estimate of the residual in the equation.
In that sense, this Note also complements our prior work~\cite{toosi:caf:20}
on grid adaptation for LES,
that used the same quantity (the divergence of the GIE)
as its error indicator,
but motivated its use from a different point-of-view 
of solution sensitivity.

\section*{Acknowledgements}

This work has been supported by 
NASA grant 80NSSC18M0148.

\bibliographystyle{elsarticle-num}
\bibliography{./references.bib}

\end{document}